\documentclass[12pt]{article}
\usepackage{fullpage}
\pdfoutput=1
\usepackage{mathrsfs}
\usepackage{amsmath}
\usepackage{amsthm}
\usepackage{amssymb}
\usepackage{graphicx}

\newcommand{\ud}{\mathrm{d}}
\newcommand{\Kn}{K\! n}

\newcommand{\MB}{\text{\tiny{MB}}}
\newcommand{\bc}{\boldsymbol{c}}
\newcommand{\bx}{\boldsymbol{x}}
\newcommand{\bu}{\boldsymbol{u}}

\DeclareMathOperator{\erf}{erf}

\title{On the Second-order Temperature Jump Coefficient of a Dilute Gas}
\author{
Gregg A. Radtke and Nicolas G. Hadjiconstantinou\footnote{Author to whom correspondence should be addressed.  Electronic mail: ngh@mit.edu}\\
Mechanical Engineering Department, Massachusetts Institute of Technology\\
Cambridge, MA 02139, U.S.A.\\
\smallskip\\
Shigeru Takata and Kazuo Aoki\\
Department of Mechanical Engineering and Science, Kyoto University\\
Kyoto 606-8501, Japan}

\date{\today}

\begin{document}

\maketitle

\begin{abstract}
We use LVDSMC simulations to calculate the second-order temperature jump coefficient for a dilute gas 
whose temperature is governed by the Poisson equation with a constant forcing term. Both the hard sphere gas and the 
BGK model of the Boltzmann equation are considered. Our results show that the temperature jump coefficient is different 
from the well known linear and steady case where the temperature is governed by the homogeneous heat conduction (Laplace) equation. 
\end{abstract}

\section{Introduction}
Slip-flow theory is a powerful tool that enables the continued use of the Navier-Stokes description as the 
characteristic flow lengthscale ($L$) approaches 
the molecular mean free path ($\lambda$) \cite{Hadjiconstantinou2006}. It can be rigorously derived
from asymptotic solution of the Boltzmann equation in the limit $\Kn=\lambda/L\ll 1$; such an analysis shows that,  
in this limit, the Navier-Stokes description remains valid in the bulk, 
but fails near the boundaries \cite{Sone2002,Sone2007}. Fortunately, the kinetic effects associated with the inhomogeneity introduced by 
the walls are only important within a layer of thickness $O(\lambda)$ near the boundaries (known as the Knudsen layer) and can be accounted for 
by a boundary-layer type of analysis where an inner kinetic solution is matched to the outer Navier-Stokes solution 
\cite{Sone2002,Sone2007}. Slip/jump 
boundary conditions and the associated {\it non-adjustable} slip coefficients emerge from this analysis as the matching condition between 
the inner and outer solution \cite{Sone2002,Sone2007}. Carrying out such an analysis to second order in $\Kn$ 
yields second-order slip/jump models \cite{Sone2002,Sone2007}, which can be very useful in a variety of cases \cite{Hadjiconstantinou2006}.

Accurate determination of slip coefficients using this rigorous procedure is quite challenging in general and becomes 
increasingly more challenging as the order of the expansion increases. Original 
studies focused on the BGK model of the Boltzmann equation \cite{Cerc,Sone1969}, for which all first-order and 
second-order coefficients are known \cite{Sone2002,Sone2007}. The first-order coefficients for the hard-sphere gas   
have also since been calculated \cite{ohwada}. However, although the form of the slip expression 
is known to second order in $\Kn$, second-order slip coefficients  for the hard-sphere gas are mostly unknown. 

As the companion paper shows \cite{companion}, the recently developed reciprocity relations by Takata \cite{Takata,Takata2} 
can be used to calculate these coefficients. An alternative approach amounts to extracting slip coefficients from 
hydrodynamic fields by comparing solutions of the Boltzmann equation with 
Navier-Stokes solutions \cite{pof2003,Hadjiconstantinou2006}. In these approaches, in addition to high accuracy 
(including low statistical uncertainty if a stochastic method is used for solving the Boltzmann equation), care needs 
to be exercised to avoid comparison of the two solutions in the Knudsen layer, where the Navier-Stokes solution is not 
equivalent to the Boltzmann solution \cite{Hadjiconstantinou2006}. This has led to a number of erroneous results in the 
past. 
  
In this paper we use this process to calculate the second-order temperature jump coefficient for a dilute gas 
when the temperature field is governed by the Poisson equation with constant forcing term. 
We calculate this coefficient using the  recently developed low-variance deviational Monte Carlo simulation method 
\cite{Homolle2007a,Homolle2007b,Radtke2009,jht2010,pof2011}, 
which is naturally suited to low-signal problems and thus allows calculations at infinitessimal temperature differences. 
The latter are necessary because finite temperature differences introduce density gradients and temperature-dependent 
transport coefficients which may alter the result.  

Our result is verified and put on a more firm theoretical footing
by the companion paper \cite{companion} which considers a mathematically equivalent time-dependent problem, thus clarifying 
why the temperature jump law and coefficient  reported here are in general different from the one obtained 
by linear {\it steady-state} analysis \cite{Sone2002}.

\section{Background}
We consider a dilute hard-sphere gas of molecular mass $m$ and molecular diameter $\sigma$, 
in contact with a planar diffusely reflecting boundary at temperature $T_\text{B}$. We also consider the BGK model of 
such a gas, with collision frequency $\tau^{-1}$. 
In the case of the hard-sphere gas, $\lambda=(\sqrt{2}\pi n_0 \sigma^2)^{-1}$, while for the BGK gas 
$\lambda=2c_0 \tau/\sqrt{\pi}$, where 
$c_0=\sqrt{2RT_0}$ is the most probable speed based on the reference temperature $T_0$, $n_0$ is a reference number density, $R=k_B/m$ is the gas constant and $k_B$ is Boltzmann's constant.

The first-order temperature jump condition at the gas-wall interface is given 
by \cite{Sone2002,Sone2007} 
\begin{equation}
\hat{T}\big\vert_\text{B} - \hat{T}_\text{B} = d_1 k \frac{\partial \hat{T}}{\partial \hat{n}}\Big\vert_\text{B}, \label{eq:Tjump1}
\end{equation}
where $\hat{T}=T/T_0$, $k=\frac{\sqrt{\pi}}{2}\Kn=\frac{\sqrt{\pi}}{2}(\lambda/L)$, $\vert_\text{B}$ denotes the boundary location, $\hat{n}$ is the unit  
(inward) normal direction and $L$ is the characteristic system length scale; the numerical constant $d_1$ obtains the {\it non-adjustable} values  
of $2.4001$ for a hard sphere gas 
and $1.30272$ for a BGK gas \cite{Sone2002}; we emphasize that these values correspond to diffusely-reflecting boundaries. 

The utility of first-order slip/jump models primarily depends 
on the amount of error that can be tolerated.  Temperature jump coefficients (both first-order and the second-order measured here) turn out 
to be larger than their velocity slip counterparts.  As a result, a second-order temperature jump correction 
becomes important at smaller Knudsen numbers. In other words, 
the first-order result (\ref{eq:Tjump1}) is typically adequate for $\Kn< 0.1$.

Asymptotic expansion to second order in $k$ \cite{Sone2002,Sone2007} for {\it linear and steady} problems  
extends (\ref{eq:Tjump1}) to the following jump condition
\begin{equation}
\hat{T}\big\vert_\text{B} - \hat{T}_\text{B} = d_1 k \frac{\partial \hat{T}}{\partial \hat{n}}\Big\vert_\text{B}  + 
d_3 k^2 \frac{\partial^2 \hat{T}}{\partial \hat{n}^2}\Big\vert_\text{B}.  \label{eq:Tjump2}
\end{equation}
This condition is valid for a quiescent gas---more precisely, a gas that is quiescent under no-slip boundary conditions; in 
the presence of gas flow, additional terms related to the flow stress need to be included \cite{Sone2002}. 
We also emphasize that according to the analysis that yields this condition, for linear and steady problems, energy conservation reduces to 
\begin{equation}
\nabla^2 \hat{T}=\frac{\partial^2 \hat{T}}{\partial \hat{x}^2}+\frac{\partial^2 \hat{T}}{\partial \hat{y}^2}+\frac{\partial^2 \hat{T}}{\partial \hat{z}^2}=0,
\label{homo}
\end{equation}
where $(\hat{x},\hat{y},\hat{z})=(x/L,y/L,z/L)$, and $L$ is the characteristic problem lengthscale.
In the special case of one-dimensional problems, equation (\ref{homo}) further reduces to  
\begin{equation}
\nabla^2 \hat{T}=\frac{\ud^2 \hat{T}}{\ud \hat{n}^2}=0,
\end{equation}
which makes the value of $d_3$ irrelevant. This is actually utilized below to calculate the slip coefficient 
due to a forcing term in the temperature equation. 

In summary, jump condition (\ref{eq:Tjump2}) is to be used when the governing equation is (\ref{homo}).
Within this approximation, $d_3$ is only known ($d_3=0$) for the special case of the BGK model \cite{Sone2002,Sone2007}. We also 
note that Deissler's result \cite{Deissler} for second-order velocity slip and temperature jump is based on approximate mean-free-path arguments 
and does not correspond to a self-consistent solution of the Boltzmann equation; as a result, it captures neither the correct form of the slip/jump  relation 
nor the correct values of the slip coefficients (e.g. compare equations (3.40)-(3.42) in \cite{Sone2007} to equations (24a) and (51) in \cite{Deissler}).

\section{Calculation of the temperature jump coefficient}
To extract the slip coefficient in a dilute gas governed by the Poisson equation with constant forcing term, 
we simulate the steady state of a one-dimensional gas layer bounded by two isothermal, 
diffuse walls at $x=\pm L/2$ and at temperature $T_0$, subject to volumetric heating at a constant rate  $\dot{Q}$.
In dimensionless form, the one-dimensional heat equation with constant volumetric heating can be written as
\begin{equation}
\nabla^2 \hat{T}=\frac{\ud^2 \hat{T}}{\ud \hat{x}^2} = -\frac{5\epsilon}{4\gamma_2k}, \label{eq:heat}
\end{equation}
where $\hat{x}=x/L$, $\gamma_2$ is a dimensionless form of the thermal conductivity---
equal to  $1.9228$ for hard spheres and unity for BGK \cite{Sone2007}---and 
\begin{equation}
\epsilon = \frac{L\, \dot{Q}}{c_0 P_0} \ll 1
\end{equation}
is the dimensionless form of the volumetric heat addition rate. Here, $P_0=n_0k_BT_0$ is a reference pressure.

The asymptotic analysis yielding (\ref{eq:Tjump2}) does not apply to the non-homogeneous equation (\ref{eq:heat}). 
A rigorous derivation 
which takes the inhomogeneous term into account by considering an equivalent unsteady problem can be found in the 
companion paper \cite{companion}, which shows that 
in a quiescent gas, in one spatial dimension, the resulting second-order slip relation is given by
\begin{equation}
\hat{T}\big\vert_\text{B}- \hat{T}_\text{B} = d_1 k \frac{\partial \hat{T}}{\partial \hat{n}}\Big\vert_\text{B} + 
d^\prime_3 k^2 \frac{\partial^2 \hat{T}}{\partial \hat{n}^2}\Big\vert_\text{B}.  \label{eq:Tjump3}
\end{equation}
We emphasize that, although the structure of the slip relation is the same as in equation (\ref{eq:Tjump2}), 
the second-order coefficient is different. It is also convenient that (\ref{eq:Tjump3}) does not contain 
$d_3$; this allows calculation of $d_3^\prime$  from volumetric heating calculations without explicit knoweldge of $d_3$.
This last feature, as well as the similarity of (\ref{eq:Tjump2}) and (\ref{eq:Tjump3}) is due to fortuitous cancellation; 
as discussed further in section \ref{disc}, under more general conditions (e.g. higher spatial dimensions),
this cancellation does not take place and terms containing both $d_3$ and $d_3^\prime$ appear.

The solution to Equation (\ref{eq:heat}) subject to boundary condition (\ref{eq:Tjump3}) is 
\begin{equation}
\hat{T} = \frac{1}{2}\frac{4\epsilon }{5\gamma_2 k} \left[ \left( \frac{1}{4}-\hat{x}^2 \right) + d_1 k - 2d_3^\prime k^2 \right] \label{eq:Tjump}.
\end{equation}
Comparison of this solution to LVDSMC simulations {\it away from the Knudsen layer} allows us to calculate the 
coefficient $d_3^\prime$. In this work, we extract the value of  $d_3^\prime$ from the slope of
\begin{equation}
\frac{5\gamma_2}{4\epsilon}\hat{T}(\hat{x}=0) - \frac{1}{8k} - \frac{d_1}{2} 
\end{equation}
as a function of $k$ for $k\rightarrow 0$.

\section{Computational method}

The Low-variance Deviational Simulation Monte Carlo (LVDSMC) method 
\cite{Homolle2007a,Homolle2007b,Radtke2009,jht2010,pof2011} efficiently simulates \cite{Wagner2008}
the Boltzmann equation 
\begin{equation}
\frac{\partial f}{\partial t} + \bc \cdot \frac{\partial f}{\partial \bx} = \left[\frac{\partial f}{\partial t}\right]_\text{coll}
\label{basicBoltzmann}
\end{equation}
written here in the absence of external body forces, 
by simulating only the \emph{deviation} $f^\ud = f - f^\MB$ from an equilibrium state $f^\MB$. 
Here, $f=f(\bx,\bc,t)$ is the single particle distribution function \cite{Sone2002}.   
This approach results in a greatly reduced level of statistical uncertainty for low signal problems compared 
to the standard DSMC \cite{Bird1994} approach and is therefore well suited to the present application. 

Volumetric heating is modeled by simulating  the equation   
\begin{equation}
\frac{\partial f}{\partial t} + \bc \cdot \frac{\partial f}{\partial \bx} = \left[\frac{\partial f}{\partial t}\right]_\text{coll}+\frac{\dot{Q}}{ P_0} \left(\frac{2}{3}\frac{c^2}{c_0^2} - 1\right) f^0
\label{heatedBoltzmann}
\end{equation}
where 
\begin{equation}
f^0  = \frac{\rho_0 }{\pi^{3/2} c_0^3}e^{-||\bc||^2/c_0^2}
\end{equation}
and $\rho_0=mn_0$ is a reference mass density. More details on the simulation of the additional term on the right hand side can be found in section \ref{collision}.

In the versions implemented here, equilibrium is described by a Maxwell-Boltzmann distribution 
\begin{equation}
f^\MB  = \frac{\rho_\MB}{\pi^{3/2} c_\MB^3} \exp\left(-\frac{||\bc-\bu_\MB||^2}{c_\MB^2}\right), \label{eq:fMB}
\end{equation}
based on local (cell-based) mass density $\rho_\MB$, velocity $\bu_\MB$, temperature $T_\MB$, and most probable velocity $c_\MB=\sqrt{2RT_\MB}$.  
Because $f^\ud$ can take positive and negative values, it is represented by signed (or deviational) particles.

As in the DSMC approach, LVDSMC solves the Boltzmann transport equation through a time-splitting approach using a 
timestep $\Delta t$. The associated advection and collision substeps are described below.

\subsection{Advection substep}
During the advection substep, particles move according to the standard DSMC procedure 
(i.e. for particle $i$, $\bx_i (t+\Delta t) = \bx_i (t) + \bc_i \Delta t$), with additional particles generated at the boundaries and cell interfaces.  
Each of these additional generation steps are implemented by drawing particles from differences of fluxal distributions.

At a stationary boundary, particles are generated by sampling from
\begin{equation}
\bc\cdot\boldsymbol{n} \left( \rho_\text{\tiny B}\phi^\text{\tiny B} - f^\MB \right) \Delta A \Delta t \ud^3 \bc,
\end{equation}
where $\Delta A$ is the surface area element at the boundary, $f^\MB$ is the equilibrium distribution in the cell 
adjacent to the boundary, and $\phi^\text{\tiny B}$ is the ``boundary distribution'' given by
\begin{equation}
\phi^\text{\tiny B}  = \frac{e^{-c^2/c^2_\text{\tiny B}}}{\pi^{3/2} c_\text{\tiny B}^3},
\end{equation}
where the $c_\text{\tiny B}=\sqrt{2RT_\text{\tiny B}}$; the ``boundary density'' $\rho_\text{\tiny B}$ is evaluated from the mass conservation statement 
\begin{equation}
\rho_\text{\tiny B} \int_{\bc\cdot\boldsymbol{n} > 0} \bc\cdot\boldsymbol{n} \,\phi^\text{\tiny B} \ud^3 \bc = -\int_{\bc\cdot\boldsymbol{n} < 0} \bc\cdot\boldsymbol{n} f^\MB \ud^3 \bc.
\end{equation}

Particles are also generated at the cell interfaces to account for the spatial discontinuities in $f^\MB$ \cite{Homolle2007b,Radtke2009}; 
they are sampled from
\begin{equation}
\bc\cdot\boldsymbol{n} \left( f^\MB_- - f^\MB_+ \right) \Delta A^\text{int} \Delta t \ud^3 \bc,
\end{equation}
where $\Delta A^\text{int}$ is the area of the interface, $f^\MB_\pm$ are the equilibrium distributions in adjacent cells, 
and $\boldsymbol{n}$ points from $f^\MB_-$ to $f^\MB_+$.

\subsection{Collision substep}
\label{collision}
The collision substep treatment is based on 
published LVDSMC implementations \cite{Radtke2009,pof2011}, suitably 
modified to include the effect of volumetric heating. We first discuss 
the BGK collision operator and the corresponding volumetric heating 
implementation; the hard-sphere case follows. Due to the small deviations from 
equilibrium, here we consider the linearized form of these collision operators; methods for 
simulating the corresponding non-linear versions can be found in \cite{jht2010,Homolle2007b,Wagner2008}.

\subsubsection{BGK model}
In the case of the BGK model, the collision operator is given by
\begin{equation}
\left[\frac{\partial f}{\partial t}\right]_\text{coll} = -\frac{f-f^\text{loc}}{\tau},
\end{equation}
where $f^\text{loc}$ is the local equilibrium distribution given by
\begin{equation}
f^\text{loc} = \frac{\rho(\bx,t)}{\left[2\pi RT(\bx,t) \right]^{3/2}} \exp \left( - \frac{||\bc-\bu(\bx,t)||^2}{2RT(\bx,t)} \right),
\end{equation}
where $\rho(\bx,t)$, $\bu(\bx,t)$ and $T(\bx,t)$ are the local mass density, flow velocity and temperature.

Using the approach of Ref. \cite{Radtke2009}, the collision step for the BGK collision operator is written as 
\begin{equation}
\left[\frac{\partial f^\ud}{\partial t}\right]_\text{coll} \Delta t = \underbrace{\frac{\Delta t}{\tau}\left[f^\text{loc} - f^\MB\right] - \Delta f^\MB}_\text{generation} - \underbrace{\frac{\Delta t}{\tau}f^\ud}_\text{deletion}, \label{eq:fMBcoll}
\end{equation}
where $\Delta f^\MB$ is a shift in the equilibrium state.  The terms above represent a source term for generating new particles, 
and a sink term for deleting existing particles.  It can be shown \cite{Radtke2009} that the generation term is eliminated for linear problems 
when the equilibrium state (for each cell) is shifted according to
\begin{equation}
\begin{bmatrix}
\rho_\MB \\
\bu_\MB \\
T_\MB
\end{bmatrix} (t+\Delta t)
= \begin{bmatrix}
\rho_\MB \\
\bu_\MB \\
T_\MB
\end{bmatrix} (t)
+\frac{\Delta t}{\tau}
\begin{bmatrix}
\rho - \rho_\MB \\
\bu - \bu_\MB \\
T - T_\MB
\end{bmatrix}(t).
\end{equation}

\noindent This results in a substantial simplification to step (\ref{eq:fMBcoll}), which reduces to the very simple 
operation of randomly deleting particles with probability $\Delta t / \tau$.

Because the above method simulates a local equilibrium $f^\MB$ that is updated in the course of the simulation, the heat 
generation term can be introduced directly (and analytically) into the algorithm using 
\begin{equation}
\dot{Q} = \rho_0\frac{\ud}{\ud t}\left(\frac{3}{2}RT_{MB}\right),
\end{equation}
which results in the following update for the temperature parameter of the equilibrium distribution
\begin{equation}
\Delta T_\MB = \frac{2\dot{Q}\Delta t}{3\rho_0 R}
\end{equation}
every timestep.

\subsubsection{Hard Sphere model}

The hard sphere collision operator is given by
\begin{equation}
\left[\frac{\partial f}{\partial t}\right]_\text{coll} = \frac{1}{m}\int\int \left(f' f'_* - ff_* \right) \frac{\sigma^2}{4} ||\bc-\bc_*|| \ud^3\bc_*\, \ud^2\boldsymbol{\Omega},
\end{equation}
where primes denote post-collision values and $\boldsymbol{\Omega}$ is the spherical angle.  The collision step for this approach
\begin{equation}
\left[\frac{\partial f^\ud}{\partial t}\right]_\text{coll} = \underbrace{\int \left[ 2K^{(1)} - K^{(2)} \right] (\bc,\bc_*)f_* \ud^3\bc}_\text{generation} - \underbrace{\nu f}_\text{deletion}
\end{equation}
is processed as a series of Markov particle generation and deletion steps as proposed by Wagner \cite{Wagner2008}; the specific algorithms employed are discussed in detail in Refs. \cite{Wagner2008,pof2011}.  In the above, 
\begin{align}
K^{(1)} (\bc,\bc_*) &= \frac{\sigma^2 \rho_\MB}{\sqrt{\pi} m c_\MB|| 
\bc-\bc_*||}\exp\left( - \frac{[(\bc-\bu_\MB)\cdot(\bc-\bc_*)]^2}{c_ 
\MB^2||\bc-\bc_*||^2} \right)\\
K^{(2)} (\bc,\bc_*) &= \frac{\pi \sigma^2}{m} ||\bc-\bc_*|| f^\MB(\bc)\\
\nu(\bc) &= \frac{\pi \sigma^2 \rho_\MB c_\MB}{m} \left[ \frac{e^{- \xi^2}}{\sqrt{\pi}} + 
\left( \xi + \frac{1}{2\xi} \right) \erf  \left( \xi \right) \right]
\end{align}
where $\xi = ||\bc-\bu_\MB||/c_\MB$.

In this approach, $f^\MB$ is not updated during the collision step because the hard-sphere simulation algorithm used here 
is based \cite{pof2011,Thesis} on the fixed global equilibrium distribution $f^0$.
However, to improve accuracy for the low values of $\Kn$ considered here,\footnote{As shown in \cite{Radtke2009}, due to the increasing importance of the local equilibrium distribution as $\Kn\rightarrow 0$, LVDSMC simulations with a variable equilibrium distribution significantly outperform their counterparts with fixed equilibrium distribution, because they can be set up to track the local equilibrium distribution and thus minimize the number of particles required for the same solution fidelity.} 
we have developed a special algorithm which uses an equilibrium distribution ($f^\MB$) 
that is not updated during the collision step but is, however,  
spatially dependent. In order to determine a suitable form of $f^{MB}$, at the early stages of the simulation, this distribution tracks the local equilibrium 
distribution  (similarly to the BGK algorithm described above) using an iterative algorithm in which  
$\rho_\MB$ and $T_\MB$ are taken from the solution at the 
previous iteration, while the velocity $\bu_\MB$ is taken to be zero. 
This process is started with $f^\MB=f^0$ and iterated until $f^\MB$ no longer changes appreciably, which usually takes less than 2 iterations.

 The uniform heat generation is implemented in this case by generating particles from the distribution
\begin{equation}
\left[ \frac{\partial f^\ud}{\partial t} \right]_{\text{heat}} =  \frac{\dot{Q}}{ P_0} \left(\frac{2}{3}\frac{c^2}{c_0^2} - 1\right) f^0. \label{eq:fdheat}
\end{equation}
Algorithms for efficiently sampling from distributions of the form  (\ref{eq:fdheat}) are described elsewhere \cite{Radtke2009,pof2011,Thesis}.

\section{Results}

Numerical simulations of the uniform heat generation problem were performed in order to extract the second-order jump 
coefficients by comparing the calculated steady centerline temperature $\hat{T}(\hat{x}=0)$ with the 
prediction of equation (\ref{eq:Tjump}) at $\hat{x}=0$.  

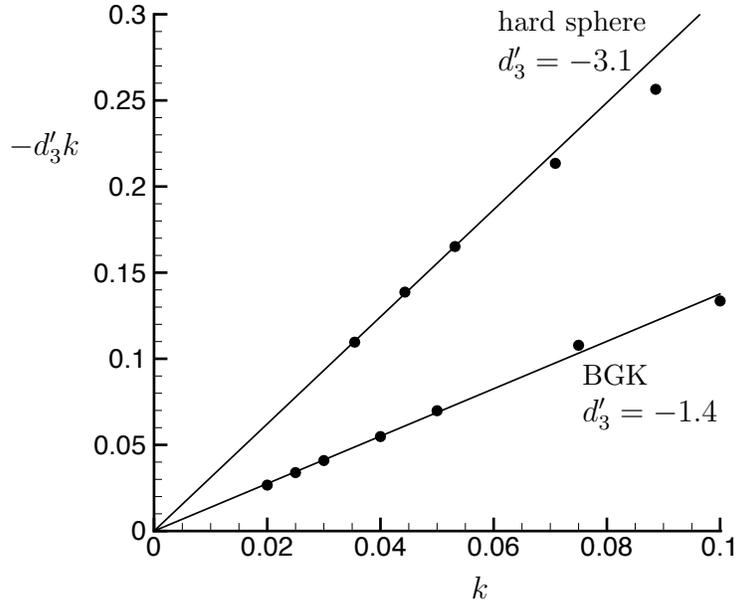
\begin{figure}
\begin{center}
\begin{picture}(280,230)(0,0)
\put(28,12){\includegraphics[width=250pt]{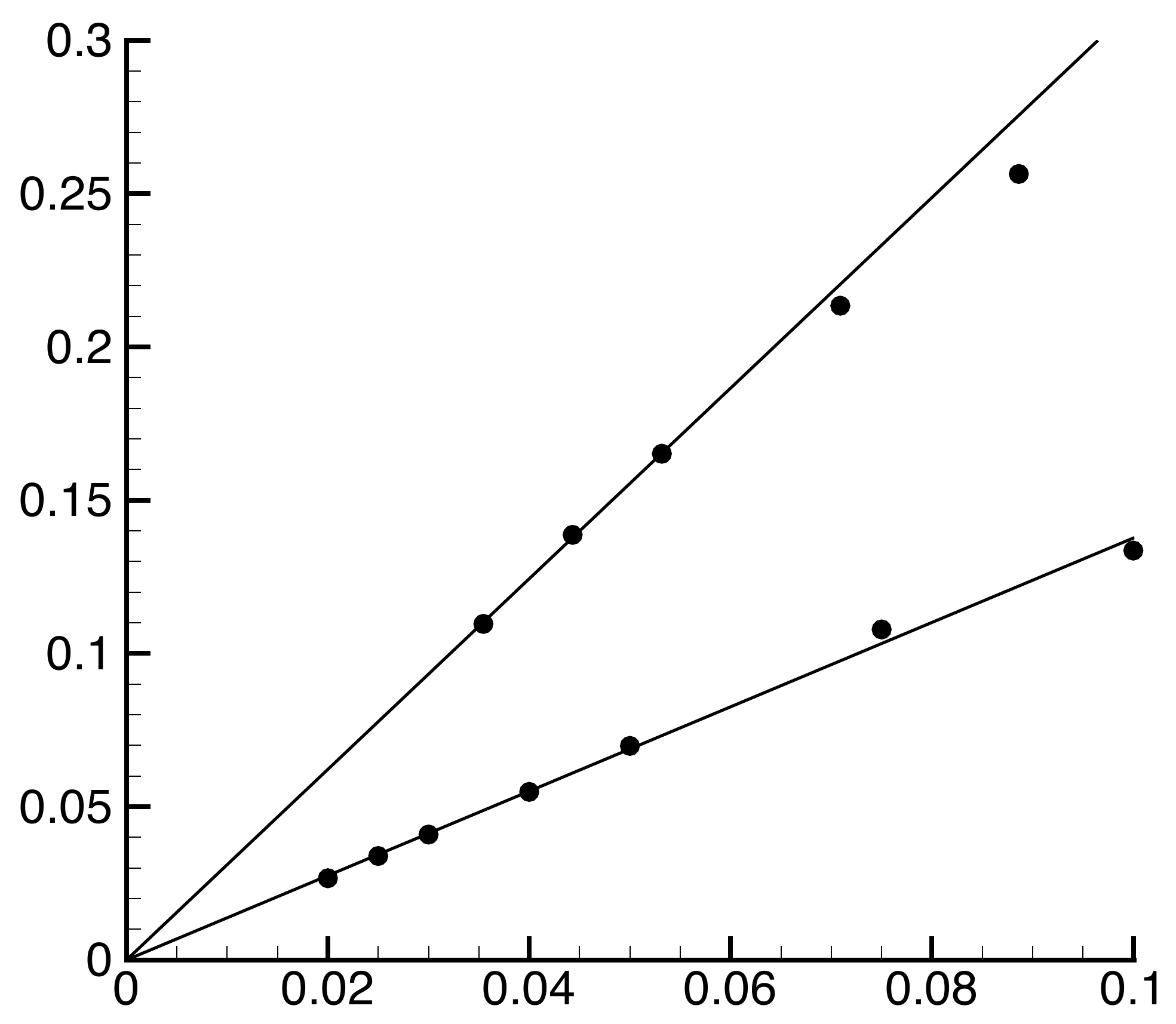}}
\put(175,0){$k$}
\put(0,168){$-d_3^\prime k$}
\put(217,81){\small BGK}
\put(217,67){$d_3^\prime=-1.4$}
\put(185,215){\small hard sphere}
\put(185,200){$d_3^\prime=-3.1$}
\end{picture}
\caption{\label{fig:d3fit} Fits used to extract the second-order jump coefficient $d_3^\prime$ for the hard sphere and 
BGK collision models.}
\end{center}
\end{figure}

Figure \ref{fig:d3fit} shows our numerical data for $-k d_3^\prime$ and a linear least squares fit passing through the origin based on the data 
for $k<0.06$, and the values $d_1=1.30272$ for BGK and $d_1=2.4001$ for the hard sphere gas \cite{Sone2007}.  
These fits yield $d_3^\prime=-1.4$ for BGK and $-3.1$ for the 
hard sphere model; the fit quality demonstrates that the leading order term is indeed $k^2$. The contribution of higher order terms starts 
to be noticable as $k$ increases. Incidentally, the complementary analysis of the companion paper \cite{companion}, based 
on a finite difference analysis of the Knudsen-layer problem of the linearized Boltzmann equation, yields 
$d_3^\prime=-1.4276$ for the BGK model and $d_3^\prime=-3.180$ for the hard-sphere model. 

\begin{figure}
\begin{center}
\begin{picture}(280,240)(0,0)
\put(30,12){\includegraphics[width=260pt]{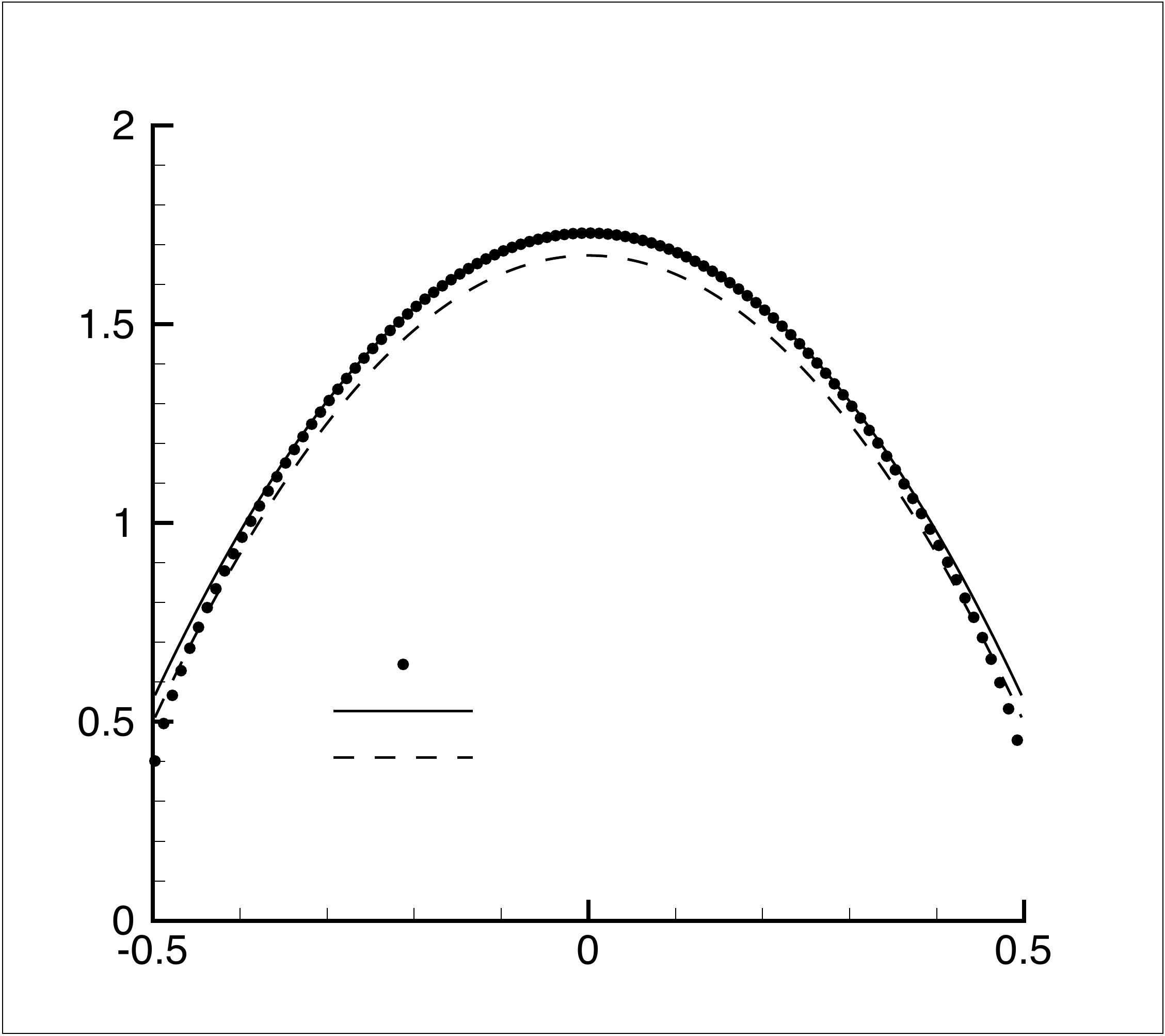}}
\put(163,0){$\hat{x}$}
\put(8,158){$\epsilon^{-1}\hat{T}$}
\put(148,94){\small simulation}
\put(148,81){\small second-order}
\put(148,68){\small first-order}
\end{picture}
\caption{\label{fig:Knp05fit} Second-order temperature jump solution (Equation (\ref{eq:Tjump})) to the uniform heat 
generation problem with Knudsen number $\Kn=0.05$; 
 simulation results (symbols) are compared to the first- (dashed line) and second-order (solid line) jump theories (equation (\ref{eq:Tjump}) with $d_3^\prime=0$ and $d_3^\prime=-3.1$, respectively).}
\end{center}
\end{figure}

\begin{figure}
\begin{center}
\begin{picture}(280,240)(0,0)
\put(30,12){\includegraphics[width=260pt]{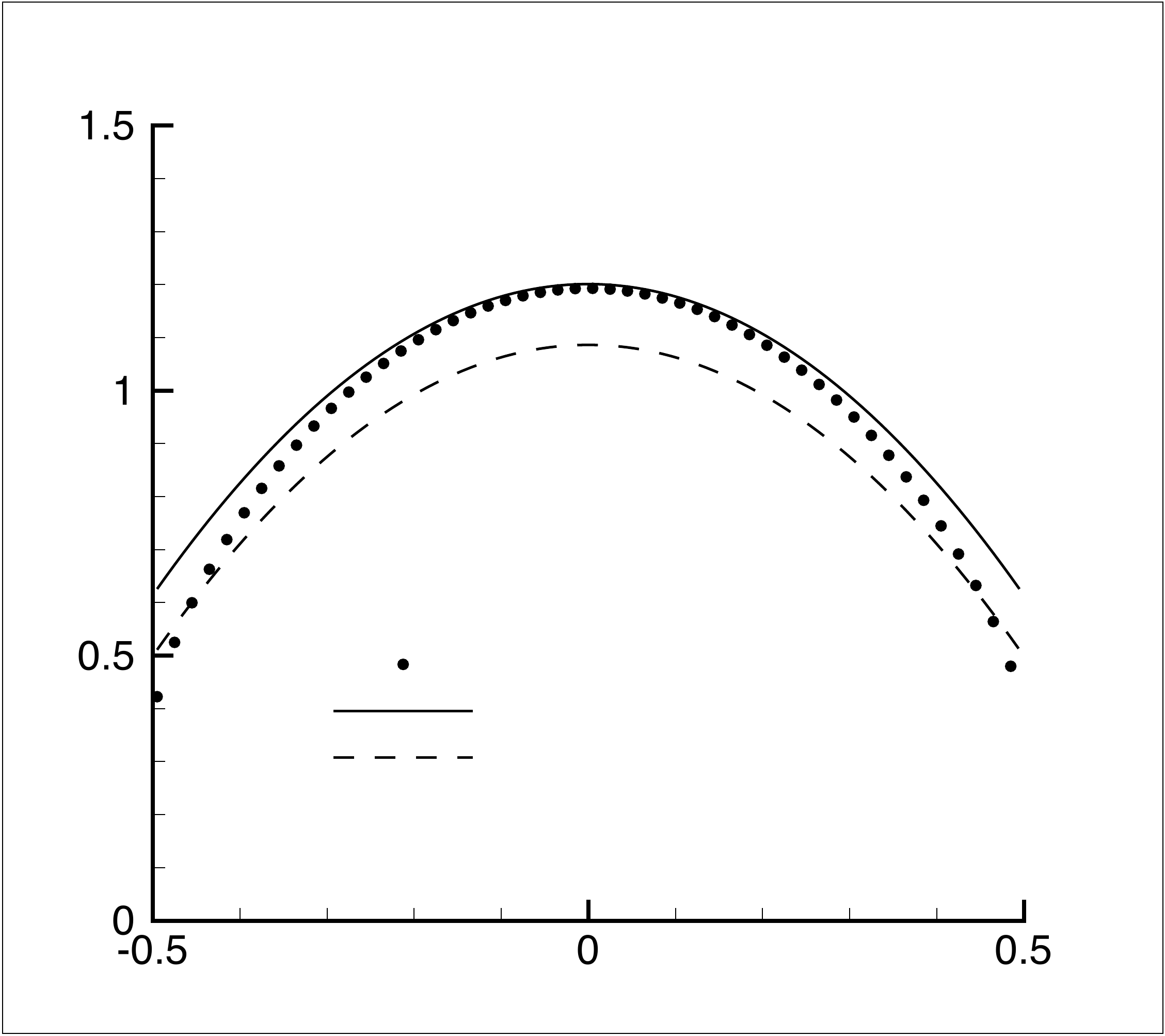}}
\put(163,0){$\hat{x}$}
\put(13,158){$\epsilon^{-1}\hat{T}$}
\put(148,94){\small simulation}
\put(148,81){\small second-order}
\put(148,68){\small first-order}
\end{picture}
\caption{\label{fig:Knp1fit} Second-order temperature jump solution (Equation (\ref{eq:Tjump})) to the uniform heat generation problem with Knudsen number $\Kn=0.1$; 
simulation results (symbols) are compared to the first-order (dashed line) and second-order (solid line) jump theories (equation (\ref{eq:Tjump}) with $d_3^\prime=0$ and $d_3^\prime=-3.1$, respectively).}
\end{center}
\end{figure}

Figure \ref{fig:Knp05fit} shows the temperature field for the hard sphere case with $\Kn=0.05$ (equivalent to $k=0.0443$) using the value obtained above 
(namely $d_3^\prime=-3.1$) demonstrating excellent agreement everywhere except in the Knudsen layer close to the boundary, as expected.  
By comparing the first- and second-order jump 
theories, it is clear that the second-order jump theory provides an improvement over the existing first-order theory, already at $\Kn=0.05$.  
For $\Kn=0.1$ (Figure \ref{fig:Knp1fit}), the error in the first order solution is quite large, while the second-order 
solution is considerably 
more accurate, provided that the existence of the Knudsen layers  for a large part of the domain is accounted for. 

\section{Discussion}
\label{disc}
Using LVDSMC simulations, we have extracted the second-order temperature jump coefficient for a hard-sphere and a BGK 
gas in the case that the Navier-Stokes-limit behavior is captured by an inhomogeneous heat conduction 
equation, such as the one appearing in the presence of constant volumetric heating. Our results have been validated 
by a companion paper which provides a deterministic calculation of the same coefficient through a rigorous 
asymptotic analysis of the Boltzmann description of a mathematically equivalent problem, namely 
that of a quescient gas confined between two parallel walls whose temperature changes linearly (increases or 
decreases) in time at a constant (and small) rate. Due to the time-dependent nature of the latter problem, 
the analysis in the companion paper goes beyond the asymptotic theory for steady problems \cite{Sone2002}; 
this also  explains why the presently calculated jump coefficient ($d_3^\prime$) is not equivalent to the one ($d_3$)  
obtained by the steady asymptotic analysis  of Ref. \cite{Sone2002}. 

Equation (\ref{eq:heat}) 
and boundary condition (\ref{eq:Tjump3}) can be generalized to two and three-dimensional steady problems as long as 
the heat generation in the gas is uniform in space and constant in time. Specifically, for a quiescent gas, 
the governing equation and boundary condition in this case become
\begin{equation}
\nabla^2 \hat{T}=-\frac{5\epsilon}{4\gamma_2 k}  
\end{equation}
and
\begin{align}
\hat{T}\big\vert_\text{B}- \hat{T}_\text{B} = (d_1+&d_5\bar{\kappa}k)  k \frac{\partial \hat{T}}{\partial \hat{n}}\Big\vert_\text{B} + 
d^\prime_3 k^2 \frac{\partial^2 \hat{T}}{\partial \hat{n}^2}\Big\vert_\text{B} \nonumber \\
& \qquad +(d_3^\prime-d_3)k^2\left(\nabla^2 \hat{T}- \frac{\partial^2 \hat{T}}{\partial \hat{n}^2}\right)\Bigg\vert_\text{B}\label{eq:Tjumpgeneral}
\end{align}
respectively.  Here $\bar{\kappa}/L$ is the mean boundary curvature and $d_5=1.82181$ for the BGK model \cite{Sone2002}; for the hard-sphere gas the value of $d_5$ is unknown.
The temperature jump coefficient $d_3$ for the hard sphere gas, 
as well as a general second-order slip 
description of unsteady problems  remain unknown and will be the subject of future work. 

\section{Acknowledgements}
This work was supported in part by the Singapore-MIT Alliance. 
NGH would like to thank KA for his hospitality  during 
his visit to Kyoto University in 2010.

\end{document}